\documentstyle[12pt]{article}
\begin{document}

\title{Biharmonic Conformal Field Theories}
\author{Franco Ferrari \\
Dipartimento di Fisica, Universit\`a di Trento,\\
38050 Povo (TN), Italy and \\
INFN, Gruppo Collegato di Trento}
\date{July 1995}
\maketitle

\vspace{-3.0in} \hfill{Preprint UTF 353} \vspace{3.5in}

\begin{abstract}
The main result of this paper is the construction of a conformally covariant
operator in
two dimensions acting on scalar fields and
containing fourth order derivatives.
In this way it is possible to derive a class of Lagrangians invariant
under conformal transformations. They define conformal field theories
satisfying equations of the biharmonic type. Two kinds of these
biharmonic field theories are distinguished, characterized by the
possibility or not of the scalar fields to transform non-trivially under
Weyl transformations. Both cases are relevant for string theory and two
dimensional gravity. The biharmonic conformal field theories provide in fact
higher order corrections to the equations of motion of
the metric and give a possibility of adding new terms to the Polyakov action.
\end{abstract}

\vfill\eject

\section{Introduction}

The main result of this paper is the construction of an operator in two
dimensions with fourth order derivatives. In this way it is possible to
derive a class of Lagrangians invariant under the Virasoro group \cite
{virasoro}. They define conformal field theories (CFT's) satisfying
equations of the biharmonic type. The biharmonic CFT's discussed here are
interesting under several points of view, from string theory \cite{gsw} to
two dimensional gravity \cite{twodimg}.
The point of view chosen in this paper starts from
the two dimensional ($2-d$) Maxwell field theory. In the Lorentz gauge, after
eliminating the longitudinal degrees of freedom, the Maxwell theory becomes
a biharmonic scalar field theory, satisfying biharmonic equations of motion $%
\bigtriangleup ^2\varphi =0$. Here $\bigtriangleup =\partial _\mu \partial
^\mu $ denotes the $2-d$ Laplacian and $\varphi $ is a scalar field. Of
course, the biharmonic equation is not conformally invariant. This is
another way to state that conformal transformations are not a symmetry of $%
2-d$ Quantum Electrodynamics. Exploiting an old theorem \cite{mitchell}, it
is however possible to make the biharmonic equation invariant under the $%
SL(2,{\rm {\bf C}})$ subgroup of the Virasoro group, allowing for point
transformations in the scalar fields $\varphi $. In this way one obtains a
curious example of $SL(2,{\rm {\bf C}})$ invariant gauge field theory,
without the need of introducing vector fields \cite{ffcqg}.

Much more difficult is the task of extending the set of symmetries of the
biharmonic equation to the whole Virasoro group. One may think for example
to construct a conformal invariant operator of the fourth order $K$ adding
to the biharmonic operator $\bigtriangleup ^2$ a suitable linear combination
of lower order operators, built out of the metric $g_{\mu \nu }$, the
covariant derivatives $\nabla _\mu $ and the Ricci tensor $R_{\mu \nu \rho
\sigma }$ \cite{gusrom}:
\[
K=\bigtriangleup ^2+aR^{\mu \nu }\nabla _\mu \nabla _\nu +bR\bigtriangleup
+cg^{\mu \nu }(\nabla _\mu R)\nabla _\nu +dR^2+
\]
\[
eR^{\mu \nu }R_{\mu \nu }+fR^{\mu \nu \rho \sigma }R_{\mu \nu \rho \sigma
}+g\bigtriangleup R
\]
Unfortunately this approach fails in two dimensions, since the coefficients $%
a-g$, chosen in such a way that $K$ becomes invariant under Weyl rescalings
of the metric $g_{\mu \nu }$, diverge.

To solve this problem, we define here a generalized Laplacian $D[A_{\mu \nu
},a_\mu]$ which transforms as follows:
\begin{equation}
D\varphi ^{\prime }=e^\phi D\varphi  \label{gentrans}
\end{equation}
under a rescaling of the fields of the kind $\varphi ^{\prime }=e^\phi
\varphi $. $D$ is constructed using the same strategy of the covariant
derivatives in gauge theories but here, besides a vector gauge field $a_\mu $%
, also a spin two gauge field $A_{\mu\nu}$ is needed in order to fulfill \ref
{gentrans}. For a conformal transformation $\zeta =\zeta (z)$, where $\zeta $
and $z$ are complex coordinates, the real function $\phi $ appearing in eq.
\ref{gentrans} is defined by $\phi =-{\rm log}\left| \frac{dz}{d\zeta }%
\right| $.

Starting from the generalized Laplacian $D$, we are able to construct a set
of interacting conformal field theories, called here of type I, with
equations of motion of the biharmonic type
\begin{equation}
D^2\varphi =\frac{\delta V(\varphi )}{\delta \varphi }  \label{eqsmot}
\end{equation}
Here $V(\varphi )$ represents an interaction potential. The above equations
of motion are invariant under the Virasoro group. However, Weyl rescalings
of the metric are a symmetry of the theory only in the free case, i.e. when $%
V(\varphi)=0$. The action of type I field theories is renormalizable if the
gauge fields $A_{\mu \nu }$ and $a_\mu $ are external. The free
energy--momentum tensor can be easily computed and it is traceless. The
problem of introducing kinetic terms also for the gauge fields without
spoiling renormalizability, locality or conformal invariance seems instead
without solution.

In a slight variation of the biharmonic CFT's presented here, called type II
for convenience, Weyl rescalings are admitted for the scalar fields, i.e. $%
\varphi ^{\prime }=e^\phi \varphi $ if the metric undergoes a Weyl
transformation of the kind $g_{\mu \nu }^{\prime }=e^{2\phi }g_{\mu \nu }$.
Clearly, the generalized Laplacian $D$ is covariant with respect to this
transformation but, as we will see, it is no longer possible to define Weyl
covariant equations of motion also at the free level. Despite of this fact,
if a conformal metric $g_{\mu \nu }=h(x)\delta _{\mu \nu }$ is chosen, the
type II biharmonic CFT's continue to remain invariant under a conformal
change of variables $\zeta =\zeta (z)$. The energy--momentum tensor is
however not traceless. The second type of biharmonic CFT's plays an
important role in $2-d$ gravity, representing possible corrections to
Liouville field theory with higher order derivatives.

The second point of view from which the biharmonic CFT's can be considered
is that of string theory and conformal field theory. Let us in fact consider
the string action in the Polyakov formulation \cite{polyakov}:
\begin{equation}
S_{str}=\int d^2x\left[ \partial _\mu X\partial ^\mu X+RX\right]
\label{polact}
\end{equation}
How much can this action be generalized, maintaining the properties of
conformal invariance and locality? This is a physically important question,
since the coupling of string theory with other CFT's yields remarkable
consequences, as it has been shown for instance in the case of the Thirring
model \cite{thirring}
discussed in ref. \cite{portom},
where a new method of spontaneous symmetry breaking
was found in this way. Unfortunately, the set of CFT's admitting a
Lagrangian formulation and which may be coupled to string theory is very
limited. The biharmonic CFT's introduced in this paper enlarge this set to a
new class of conformal field theories. Some interesting possibilities of
letting both type I and type II CFT's interact with string theory will be
given below.

A third and final point of view which will be only briefly discussed here is
that of $2-d$ gravity. In the second type of biharmonic CFT's, where the
scalar fields $\varphi $ are allowed to rescale according to a Weyl
transformation of the metric explained above, it is possible to perform the
following identification: $\varphi =\left| g\right| ^{-\frac 14}$, where $g$
is the determinant of the metric. Thus the type II biharmonic CFT's provide
local equations of motion in $\left| g\right| ^{-\frac 14}$ and consequently
in the metric. As mentioned before, the type II equations of motion are no
longer Weyl invariant. Therefore, conversely to what happens in string
theory, also the conformal factor of the metric can be determined adding
biharmonic CFT's to the Polyakov action \ref{polact}. With respect to the
Liouville field theory, the biharmonic corrections to $2-d$ gravity have the
advantage of providing simple equations, e.g. in the free case, in which the
potential $V\left( \varphi \right) $ is set to zero.

In the following, the biharmonic CFT's will be treated in details, both in
the Weyl invariant and Weyl noninvariant versions. Possible ways to couple
them to string theory will be discussed, together with their interpretation
as generalizations of the action for $2-d$ gravity containing higher order
derivatives.

\section{Biharmonic Conformal Field Theories}

Let us start by considering the action of $2-d$ Quantum Electrodynamics
(QED).
In complex coordinates $z=x_1+ix_2$, $\overline{z}%
=x_1-ix_2$, we have:
\begin{equation}
S_{Maxwell}=\int d^2zF_{z\overline{z}}^2  \label{complmax}
\end{equation}

with $F_{z\overline{z}}=\partial _zA_{\overline{z}}-\partial _{\overline{z}%
}A_z$. Complex indices will be denoted with the first letters of the Greek
alphabet, for instance $\alpha =z,\overline{z}$. Exploiting the Hodge
decomposition for the gauge fields $A_\alpha $:
\begin{equation}
A_z=\partial _z\chi +\partial _z\varphi\qquad\qquad\qquad
A_{\overline{z}}=\partial _{\overline{z}}\chi -\partial _{\overline{z}%
}\varphi  \label{hodgedec}
\end{equation}

one obtains:
\begin{equation}
S_{Maxwell}=\int d^2z(\bigtriangleup \varphi )^2  \label{biharmax}
\end{equation}
\noindent
Thus the QED action is not conformally invariant in $%
2-d$. To see this more in details, let us consider the biharmonic equations
of motion coming from \ref{biharmax}:
\begin{equation}
\bigtriangleup (\bigtriangleup \varphi )=0  \label{biharmeq}
\end{equation}

\noindent Performing a conformal change of variables $\zeta =\zeta (z)$, the
Laplacian transforms as follows: $4\partial _z\partial _{\overline{z}}=\frac
4{J^2}\partial _\zeta \partial _{\overline{\zeta }}$, where $J=\left| \frac{%
dz}{d\zeta }\right| $. Therefore, the biharmonic equation \ref{biharmeq}
becomes in the new coordinates:
\[
\frac 1{J^2}\partial _\zeta \partial _{\overline{\zeta }}\left( \frac
1{J^2}\partial _\zeta \partial _{\overline{\zeta }}\varphi \right) =0
\]
and it is clearly not conformally invariant unless $J$ is a constant.

One way to extend the set of transformations leaving the biharmonic equation
invariant is to allow for point transformations in the field $\varphi $ of
the kind \cite{mitchell}:
\begin{equation}
\varphi \left( z,\overline{z}\right) =J\varphi ^{\prime }\left( \zeta ,%
\overline{\zeta }\right)  \label{conftrans}
\end{equation}
If $\varphi$ obeys the above requirement, the action \ref{biharmax} becomes
invariant under the M\"obius group of transformations $\zeta =\frac{az+b}{%
cz+d}$, with $a,b,c,d$ being arbitrary complex numbers satisfying the
condition $ad-bc\neq 0$. As pointed out in ref. \cite{ffcqg}, this is a
remarkable theory, having an SL(2,{\bf C}) group of local symmetries and no
gauge fields. On a sphere, the gauge group coincides with the SL(2,{\bf C})
group of automorphisms admitted on a genus zero Riemann surface.
Other symmetries of the biharmonic equations have been obtained in
\cite{bluman}.
In order to
obtain invariance under the whole set of conformal transformations, however,
the introduction of gauge fields is necessary.

To this purpose, we define a new Laplacian $D_{z\overline{z}}$, requiring
that
\begin{equation}
D_{z\overline{z}}\varphi ^{\prime }(z,\overline{z})=e^{\phi (z,\overline{z}%
)}D_{z\overline{z}}\varphi (z,\overline{z})  \label{weylcov}
\end{equation}
if $\varphi $ is rescaled by an arbitrary real factor:
\begin{equation}
\varphi ^{\prime }\left( z,\overline{z}\right) =e^{\phi (z,\overline{z}%
)}\varphi \left( z,\overline{z}\right)  \label{weylt}
\end{equation}

To fulfill requirement \ref{weylcov}, let us try the following ansatz:
\begin{equation}
D_{z\overline{z}}\varphi =\left[ \partial _z\partial _{\overline{z}}+aA_{z%
\overline{z}}+b\left( a_z\partial _{\overline{z}}+a_{\overline{z}}\partial
_z\right) +c\left( a_za_{\overline{z}}\right) \right] \varphi
\label{newlapl}
\end{equation}
where $A_{z\overline{z}}$ and $a_\alpha $ are gauge fields undergoing
suitable transformations in order to reabsorb all the unwanted terms in $%
\phi \left( z,\overline{z}\right) $ which arise in the rhs of eq. \ref
{weylcov} after exploiting the derivatives $\partial _z\partial _{\overline{z%
}}$. In the case of a conformal change of variables $\zeta =\zeta (z)$, the
scalar field $\varphi $ transforms as in eq. \ref{conftrans}. Therefore,
putting $\phi (z,\overline{z})=-\log J$, with $J=\left| \frac{dz}{d\zeta }%
\right| $ in eqs. \ref{weylcov} and \ref{weylt}, it is easy to realize that
the new Laplacian $D_{z\overline{z}}$ is covariant also under conformal
transformations:
\begin{equation}
D_{\zeta \overline{\zeta }}\left[ A_{\zeta \overline{\zeta }},a_\zeta ,a_{%
\overline{\zeta }}\right] \varphi ^{\prime }\left( \zeta ,\overline{\zeta }%
\right) =JD_{z\overline{z}}\left[ A_{z\overline{z}},a_{\overline{z}}\right]
\varphi \left( z,\overline{z}\right)  \label{confcov}
\end{equation}
Eqs. \ref{weylcov} and \ref{confcov} are verified performing in eq. \ref
{newlapl} the substitutions:
\begin{equation}
a=-1\qquad b=-1\qquad c=1  \label{coeff}
\end{equation}
Moreover, the fields $A_{z\overline{z}}$ and $a_\alpha $ should obey the
following transformations:
\begin{equation}
A_{\zeta \overline{\zeta }}d\zeta d\overline{\zeta }=\left( A_{z\overline{z}%
}+\partial _z\partial _{\overline{z}}\phi \left( z,\overline{z}\right)
\right) dzd\overline{z}  \label{gaugetone}
\end{equation}
\begin{equation}
a_\zeta d\zeta +a_{\overline{\zeta }}d\overline{\zeta }=\left( a_z+\partial
_z\phi \left( z,\overline{z}\right) \right) dz+\left( a_{\overline{z}%
}+\partial _{\overline{z}}\phi \left( z,\overline{z}\right) \right) d%
\overline{z}  \label{gaugetwo}
\end{equation}
Starting from the modified Laplacian \ref{newlapl}, we can easily build the
free action of type I biharmonic CFT's:
\begin{equation}
S_{conf}=\int d^2z\left( D_{z\overline{z}}\varphi \right) ^2
\label{sconfinv}
\end{equation}
Apparently, the above Lagrangian density
\[
L_{zz\overline{z}\overline{z}}=\left( D_{z\overline{z}}\varphi \right) ^2
\]
has the wrong tensorial properties. Proper Lagrangians in complex $2-d$
coordinates should be in fact $(1,1)$ tensors $T_{z\overline{z}}$ with one
complex and a complex conjugate indices. For instance, the Lagrangian of
massless scalar fields $X(z,\overline{z})$ is given by $L_{z\overline{z}%
}=\partial _zX\partial _{\overline{z}}X$. However, due to eq. \ref{conftrans}%
, the scalar fields $\varphi $ behave as $\varphi \sim \frac 1{\sqrt{g_{z%
\overline{z}}}}$, i.e. they may be regarded as $\left( \frac 12,\frac
12\right) $ differentials. This fact assures the covariance of $S_{conf}$
also under general diffeomorphisms, as we will see below. Let us notice that
if we insist in the interpretation of $\varphi$ as a spinor, no ambiguity
with the spin structures arises, because $\varphi $ transforms as an element
of $\left( \frac 12,0\right) \otimes \left( 0,\frac 12\right) $. As a
consequence, the multivaluedness of the $\left( \frac 12,0\right) $
component induced by the presence of an hypothetical spin structure cancels
against the phase of the $\left( 0,\frac 12\right) $ component. Thus $%
S_{conf}$ describes a well defined conformal field theory with higher order
derivatives. As in string theory, the dependence on the metric disappears
once conformal coordinates are chosen. In $2-d$ this is a great advantage,
since every Riemannian manifold is conformally flat. Let us now discuss
possible kinetic terms for the gauge fields $A_{z\overline{z}}$ and $%
a_\alpha $. A simple way to introduce a dynamics for the gauge fields is
suggested by the transformation rules \ref{gaugetone}--\ref{gaugetwo} and
consists in rewriting $A_{z\overline{z}}$ and $a_\alpha $ in terms of two
other scalar fields $A$ and $a$:
\[
A_{z\overline{z}}=\partial _z\partial _{\overline{z}}A
\]
\[
a_z=\partial _za\qquad \qquad a_{\overline{z}}=\partial _{\overline{z}}a
\]
For consistency with eqs. \ref{gaugetone}--\ref{gaugetwo}, $A$ and $a$
should transform as follows under a conformal change of variables:
\[
A^{\prime }=A+\phi (z,\overline{z})\qquad \qquad a^{\prime }=a+\phi (z,%
\overline{z})
\]
It is now easy to define a conformally invariant kinetic term for $A$ and $a$%
:
\[
S_{kin}=\int d^2z\partial _z(A-a)\partial _{\overline{z}}(A-a)
\]
In this way, however, the action $S=S_{conf}+S_{kin}$ becomes no longer
renormalizable. Other attempts to build local kinetic terms for the gauge
fields either lead to non-renormalizable actions or break the conformal
symmetry. Hence, we will assume in the following that the fields $A_{z%
\overline{z}}$ and $a_\alpha $ are external fields. This is not a serious
limitation. Effective Lagrangians without the presence of those gauge fields
can be always constructed solving the classical equations of motion in $A_{z%
\overline{z}}$ and $a_\alpha $. Alternatively, one may integrate out the
external fields in the path integral using suitable weights in order to
assure convergence.

At this point, we compute the free energy momentum tensor of the conformal
field theories of type I defined by the action $S_{conf}$. Using a general,
not conformally flat metric, we obtain in real coordinates the following
expression for the functional $S_{conf}$:
\begin{equation}
S_{conf}=\int d^2x\,g\left[ \bigtriangleup \varphi -g^{\mu \nu }\left(
A_{\mu \nu }+a_\mu \partial _\nu -a_\mu a_\nu\right) \varphi \right] ^2
\label{sconfull}
\end{equation}
where $g=\left|\det g^{\mu \nu }\right| $ and $\bigtriangleup \varphi =\frac
1{\sqrt{g}}\partial _\mu \left( \sqrt{g}g^{\mu \nu }\partial _\nu \varphi
\right) $. In eq. \ref{sconfull} the determinant $g$ of the metric and not
its square root appears. The reason is the presence of the field $\varphi $
with the transformation rule \ref{conftrans}, suitably generalized to the
case of general diffeomorphisms $y^\mu=y^\mu(x)$ putting $J={\rm det}\left|%
\frac{dx^\mu} {dy^\nu}\right|$. Clearly the action \ref{sconfull} is
invariant under general diffeomorphisms and also under a Weyl rescaling of
the metric $g^{\prime}_{\mu\nu}=e^{\phi(x)}g_{\mu\nu}$. Setting $l_{\mu \nu
}\varphi =-\left( A_{\mu \nu }+a_\mu\partial _\nu -a_\mu a_\nu\right)
\varphi $, the energy momentum tensor $(T_{conf})_{\kappa \lambda }\equiv
\frac{\delta S_{conf}}{\delta g_{\mu \nu }}$ can be written as follows:
\[
\left( T_{conf}\right) _{\kappa \lambda }=2\sqrt{g}\left( \bigtriangleup
\varphi +g^{\mu \nu }l_{\mu \nu }\varphi \right) \cdot
\]
\begin{equation}
\left[ \partial _\mu \left( \frac 12\sqrt{g}g^{\kappa \lambda }g^{\mu \nu
}\partial _\nu \varphi -\sqrt{g}g^{\mu \kappa }g^{\nu \lambda }\partial _\nu
\varphi \right) +\left( \frac 12\sqrt{g}g^{\kappa \lambda }g^{\mu \nu }-%
\sqrt{g}g^{\mu \kappa }g^{\nu \lambda }\right) l_{\mu \nu }\varphi \right]
\label{tconf}
\end{equation}
{}From eq. \ref{tconf} it turns out that $(T_{conf})_{\kappa\lambda}$ is
traceless, as it should be in a conformal field theory.

{}From the transformation law \ref{conftrans}, one can also give to the field
$%
\varphi $ an interpretation as the inverse of the fourth root of the metric.
Let us discuss this case, performing the identification
\begin{equation}
\varphi =g^{-\frac 14}  \label{intone}
\end{equation}
Since $g$ is a semipositive definite quantity, (we allow also for degenerate
metrics), there is no ambiguity in taking its fourth root. Moreover, as
explained before, all the problems with the spin structures are absent,
despite of the fact that now $\varphi $ has spinorial indices. Substituting
eq. \ref{intone} in eq.\ref{sconfinv}, $S_{conf}$ becomes
\begin{equation}
S_{conf}=\int d^2 z\left[D_{z\overline{z}}\enskip
g^{-\frac 14}\right]^2\label{hocorr}
\end{equation}
and it is local in $g^{-\frac 14}$. Thus $S_{conf}$
may be considered in this case as a
possible correction of the $2-d$ gravity action, characterized by the
presence of higher order derivatives. Remarkably, after the identification
\ref{intone}, $S_{conf}$ remains conformally invariant, but clearly the
invariance under Weyl rescalings of the metric is lost. Therefore, contrary
to the Polyakov formulation of string theory, the equations of motion coming
from $S_{conf}$ are also able to determine the conformal factor of the
metric. In principle, we can define an action for $\varphi =(g)^{-\frac 14}$
that satisfies both Weyl and conformal symmetries:
\begin{equation}
S_{conf}^{\prime }=\int d^2x\enskip g^{\frac 14}
\left[\left( \bigtriangleup -g^{\mu \nu }l_{\mu \nu
}\right)\enskip g^{\frac 14}\right]\label{intwstr}
\end{equation}
In a conformally flat metric:
\[
S_{conf}^{\prime }=\int d^2z\enskip g^{-\frac 14}\left( D_{z\overline{z}%
}\enskip g^{\frac 14}\right)
\]
As we see, $S_{conf}^{\prime }$ is unfortunately non--local
in the field $\varphi =(g)^{-\frac
14}$ and therefore we will neglect it.
In general, for a field $\varphi $ transforming as $(g)^{-\frac 14}$,
we will have that the free action of type II biharmonic CFT's coincides with
$S_{conf}$.

Self-interactions of the metric field can be added for both
type I and type II theories introducing the following
functional:

\begin{equation}
S_{int}=\int d^2zV_n(\varphi )  \label{sint}
\end{equation}
Substituting
\[
G_{z\overline{z}}=A_{z\overline{z}}-\frac 12\left( \partial _za_{\overline{z}%
}+\partial _{\overline{z}}a_z\right)
\]
the most general possible potential $V_n(\varphi )$ is given by:
\begin{equation}
V_n(\varphi )=a_{-1}\varphi ^{-2}+a_0G_{z\overline{z}}+a_1G_{z\overline{z}%
}^2\varphi ^2+a_2G_{z\overline{z}}^3\varphi ^4+\ldots +a_nG_{z\overline{z}%
}^n\varphi ^{2n}  \label{confpot}
\end{equation}
with $a_1\ldots a_n$ denoting a set of real coupling constants. Eqs. \ref
{sint}--\ref{confpot} describe a well defined conformal action, invariant
under the transformations \ref{conftrans}, \ref{confcov} and \ref{gaugetone}%
--\ref{gaugetwo}. However, in both cases of type I and II biharmonic CFT's,
the action $S_{int}$ is not invariant for Weyl rescalings. The above
potential is interesting also because it yields several possibilities of
coupling the biharmonic CFT's with the basic fields of string theory, i.e.
the scalar fields $X$, the metric $g_{\mu \nu }$ and, in the super case ,
the spin currents of the kind $J_{\theta \overline{\theta }}=\overline{\psi }%
_\theta \psi _{\overline{\theta }}$.

\section{Conclusions}

Concluding, let us briefly summarize our results. Starting from the new
Laplacian defined by eqs. \ref{weylcov}, \ref{newlapl}, \ref{coeff} and \ref
{gaugetone}--\ref{gaugetwo}, we are able to construct conformal field
theories with biharmonic equations of motion. Type I
biharmonic CFT's are characterized by the free action \ref{sconfinv}. After
a conformal transformation, the scalar fields $\varphi $ undergo the point
transformation defined in eq. \ref{weylt}. The type I free biharmonic CFT is
both Weyl and conformal invariant. Its energy--momentum tensor is traceless.
The Weyl invariance is however lost if a self--interacting Lagrangian is
added (eqs. \ref{sint}--\ref{confpot}).

In type II biharmonic CFT's $\varphi $ is allowed to rescale under Weyl
transformations. The action is again provided by eq. \ref{sconfinv}, but in
this case the Weyl invariance is lost. It is however possible to define an
action with second order derivatives which is both invariant under Weyl and
conformal transformations (eq. \ref{intwstr}).
Unfortunately, this action is non-local in the scalar fields.
After the identification \ref
{intone}, the type II\ biharmonic CFT's become particularly
relevant in $2-d$ gravity and
string theory. Also for type II theories one can add to the free action
an interacting term $S_{int}$ of the kind given in eqs. \ref{sint}--%
\ref{confpot}.

Finally, many interesting issues could not be treated here because
they are outside of the aims of this letter. For instance the relation, if
it exists, between the biharmonic CFT's and the usual ones
\cite{bpz}. Also the problem
of their quantization and several aspects connected to possible consequences
in string theory could not be fully developed due to the lack of space. We
hope however to have drawn with this letter the attention to the interesting
class of biharmonic conformal field theories. They provide in fact local and
simple equations of motion for the metric in two dimensions (see e.g.
eq. \ref{hocorr}) and can be
easily coupled with the fields entering superstring theory. In this way
it seems possible to extend the Polyakov action introducing new terms and
new free parameters. Hopefully this will provide more degrees of freedom, that
will at least be sufficient to settle some of the remaining problems of
string theories.

\section{Acknowledgements}
It is a pleasure to thank Guido Cognola for useful discussion and for
pointing out ref. \cite{gusrom}.

\end{document}